\documentclass[usenatbib]{mnras}

\usepackage{bm}
\usepackage[varg]{txfonts}
\usepackage{hyperref}
\usepackage{graphicx}
\newcommand{\te}{t_{\rm E}}

\newcommand{\pie}{\pi_{\rm E}}
\newcommand{\pien}{\pi_{\rm E,N}}
\newcommand{\piee}{\pi_{\rm E,E}}
\newcommand{\bpie}{\bm{\pi}_{\rm E}}

\newcommand{\thetae}{\theta_{\rm E}}
\newcommand{\thetas}{\theta_{\rm \star}}

\newcommand{\ds}{D_{\rm S}}

\newcommand{\delcs}{\Delta \chi^{2}}

\title[Jupiter-mass planet orbiting a dwarf star]{OGLE-2018-BLG-1428Lb: a Jupiter-mass planet beyond the snow line of a dwarf star}

\author[Kim Y.-H. et al.]{
           Yun-Hak Kim,$^{1,2}$
            Sun-Ju Chung,$^{1,2}$\thanks{Corresponding author}\thanks{E-mail: sjchung@kasi.re.kr}
            Andrej Udalski,$^{3}$
            Andrew Gould,$^{4,5}$
            Michael D. Albrow,$^{6}$
            Youn Kil Jung,$^{1}$
 \newauthor Kyu-Ha Hwang,$^{1}$
            Cheongho Han,$^{7}$
            Yoon-Hyun Ryu,$^{1}$
             In-Gu Shin,$^{1}$
             Yossi Shvartzvald,$^{8}$
             Jennifer C. Yee,$^{9}$
 \newauthor  Weicheng Zang,$^{10}$          
            Sang-Mok Cha,$^{1,11}$
             Dong-Jin Kim,$^{1}$
             Hyoun-Woo Kim,$^{1,12}$
             Seung-Lee Kim,$^{1,2}$
             Chung-Uk Lee,$^{1}$             
\newauthor     Dong-Joo Lee,$^{1}$        
             Yongseok Lee,$^{1,11}$
             Byeong-Gon Park,$^{1}$
             and Richard W. Pogge$^{4}$
             (KMTNet Collaboration)
\newauthor   Przemek Mr{\'o}z,$^{3,13}$
              Radek Poleski,$^{4}$
              Marcin Wrona,$^{3}$
              Patryk Iwanek,$^{3}$
              Micha\l{} K. Szyma{\'n}ski,$^{3}$
              Jan Skowron,$^{3}$
  \newauthor  Igor Soszy{\'n}ski,$^{3}$
              Szymon Koz{\l}owski,$^{3}$
              Pawe\l{} Pietrukowicz,$^{3}$
              Krzysztof Ulaczyk$^{3,14}$
              and Krzysztof Rybicki$^{3}$
\newauthor              (The OGLE collaboration)\\        
 \\           
$^{1}$Korea Astronomy and Space Science Institute, 776 Daedeokdae-ro, Yuseong-Gu, Daejeon 34055, Republic of Korea\\
$^{2}$University of Science and Technology, Korea, (UST), 217 Gajeong-ro, Yuseong-gu, Daejeon 34113, Republic of Korea\\
$^{3}$Warsaw University Observatory, AI.~Ujazdowskie~4, 00-478~Warszawa, Poland\\
$^{4}$Department of Astronomy, Ohio State University, 140 W. 18th Avenue, Columbus, OH 43210, USA\\
$^{5}$Max-Planck-Institute for Astronomy, K{\"o}nigstuhl 17, D-69117 Heidelberg, Germany\\
$^{6}$Department of Physics and Astronomy, University of Canterbury, Private Bag 4800 Christchurch, New Zealand\\
$^{7}$Department of Physics, Chungbuk National University, Cheongju 361-763, Republic of  Korea\\
$^{8}$Department of Particle Physics and Astrophysics, Weizmann Institute of Science, Rehovot 76100, Israel\\
$^{9}$Center for Astrophysics $I$ Harvard \& Smithsonian, 60 Garden Street, Cambridge, MA 02138, USA\\
$^{10}$Department of Astronomy and Tsinghua Centre for Astrophysics, Tsinghua University, Beijing 100084, China\\
$^{11}$School of Space Research, Kyung Hee University, Giheung-gu, Yongin, Gyeonggi-do, 17104, Republic of Korea\\
$^{12}$Department of Astronomy, Chungbuk National University, Cheongju 361-763, Republic of  Korea\\
$^{13}$Division of Physics, Mathematics, and Astronomy, California Institute of Technology, Pasadena, CA 91125, USA\\
$^{14}$Department of Physics, University of Warwick, Gibbet Hill Road, Coventry, CV4 7AL, UK
 }

\date{accepted XXX. Received YYY; in original form ZZZ}
\pubyear{2020}

\begin{document}
\label{firstpage}
\pagerange{\pageref{firstpage}--\pageref{lastpage}}
\maketitle

\begin{abstract}
We present the analysis of the microlensing event OGLE-2018-BLG-1428, which has a short-duration ($\sim 1$ day) caustic-crossing anomaly. The event was caused by a planetary lens system with planet/host mass ratio $q=1.7\times10^{-3}$. Thanks to the detection of the caustic-crossing anomaly, the finite source effect was well measured, but the microlens parallax was not constrained due to the relatively short timescale ($\te=24$ days). From a Bayesian analysis, we find that the host star is a dwarf star $M_{\rm host}=0.43^{+0.33}_{-0.22} \ M_{\odot}$ at a distance $D_{\rm L}=6.22^{+1.03}_{-1.51}\ {\rm kpc}$ and the planet is a Jovian-mass planet $M_{\rm p}=0.77^{+0.77}_{-0.53} \ M_{\rm J}$ with a projected separation $a_{\perp}=3.30^{+0.59}_{-0.83}\ {\rm au}$. The planet orbits beyond the snow line of the host star. Considering the relative lens-source proper motion of $\mu_{\rm rel} = 5.58 \pm 0.38\  \rm mas\ yr^{-1}$, the lens can be resolved by adaptive optics with a 30m telescope in the future.
\end{abstract}

\begin{keywords}
gravitational lensing: micro - planets and satellites: detection
\end{keywords}

\section{Introduction}

The core accretion model proposes that gas giant planets originated beyond the snow line of their host stars, such as Jupiter and Saturn in the Solar system \citep{mizuno1980, pollack1996, inaba2003}.
This model predicts that it takes $\sim 3\, \rm Myr$ to form gas giant planets at $5\, \rm AU$ around Sun-like stars, while for low-mass stars it does not seem to be able to form such planets, because the lifetime of the proto-planetary disk for low-mass stars is not long enough to form such giant planets \citep{ida2004,laughlin2004,boss2006}.
For example, the formation time of gas giant planets for a $0.4\, M_{\odot}$ dwarf is $\gtrsim 10\, \rm Myr$, whereas its disk lifetime is $< 10\, \rm Myr$ \citep{boss2006}.
On the other hand, the disk instability model is thought to be more likely to form gas giants around beyond the snow line of M dwarfs \citep{boss2006}.
Actual detections of Jupiter-mass planets orbiting low-mass stars (\citealp{boss2006} and references therein) are consistent with the disk instability model.
Hence, these two planet formation models may actually complement one another, although this remains uncertain.

Currently, the majority of host stars with exoplanets are Sun-like stars, and their planets are mostly located inside the snow line.
Those planets have been mostly discovered by the radial-velocity and transit methods.
However, most of stars in the Galaxy are low-mass M dwarf stars, which are difficult to observe with the two methods.
On the other hand, the microlensing method typically detects low-mass M dwarfs hosting planets located beyond the snow line.
This is because the microlensing depends only the mass of objects, not the light.
Therefore, microlensing provides very important samples to constrain planet formation models including the core-accretion and gravitational instability models.

However, a majority of masses of microlensing planets were not directly measured but estimated from a Bayesian analysis, which assumes that the planet hosting probability is independent of the host star mass \citep{vandorou2020,bhattacharya2020}.
The lens masses estimated from the Bayesian analysis can be confirmed from high-resolution follow-up observations.
This is because the lens and source stars are typically separated each other within $\sim 10$ years after the peak time of event, thus making it possible to discriminate the two stars.
Until now, the masses of 18 planetary lens systems (e.g., \citealp{bennett2006, bennett2015, batista2015, fukui2015, vandorou2020, bhattacharya2020}) have been measured from high-resolution follow-up observations with Keck, VLT, Subaru, or HST.

In addition, the masses of lens systems can be directly measured from the measurement of two parameters of angular Einstein radius ($\thetae$) and microlens parallax ($\pie$).
However, it is usually hard to measure the two parameters.
This is because $\thetae$ can be measured from events with high-magnification or caustic-crossing features, while $\pie$ can be measured from the detection of the distortions induced by the orbital motion of Earth on a standard microlensing light curve \citep{gould1992}.
In general, the measurement of the microlens parallax is limited to events with long timescale $\te \gtrsim 60\, \rm days$ or large $\pie$ to detect the light-curve distortion induced by the orbital motion of Earth.
This means that for short timescale events induced by low-mass objects (e.g., M dwarfs or brown dwarfs), it is difficult to measure the microlens parallax.
The microlensing parallax measurement was first reported \footnote{The first event with a microlens parallax measurement was MACHO-LMC-5, which was discovered in 1993 \citep{alcock1997}, but the parallax measurement was only reported eight years after the event \citep{alcock2001}.} in 1995 \citep{alcock1995}, and it was due to a long timescale of the event, $\te=110\,\rm days$.
For the measurement of the microlens parallax for all events, it is required a simultaneous observation of an event from Earth and a satellite \citep{refsdal1966, gould1994}.
Then, the microlens parallax is measured from the difference in the light curves as seen from the two observatories \citep{refsdal1966, gould1994}.
Over 900 events so far have been detected from ground-based observations and the \textit{Spitzer} satellite, which is for studying the Galactic distribution of planets (\citealp{zhu2017} and the references therein).
Also, the \textit{Nancy Grace Roman} (\textit{Roman}, formerly \textit{WFIRST}) satellite will be launched in near future \citep{spergel2015}.
With this satellite, it is expected to detect $\sim 1400$ bound exoplanets \citep{penny2019} and $\sim 250$ free-floating planets \citep{johnson2020}.
Hence, the masses of over 1000 planetary systems can be measured from the \textit{Roman} together with ground-based observations, such as the Korea Microlensing Telescope Network (KMTNet; \citealp{kim2016}).
However, we note that the main mass measurement method for the \textit{Roman} Galactic Exoplanet Survey will be the detection of the exoplanet host stars in the \textit{Roman} imaging data.
The microlensing parallax between the Earth and \textit{Roman} will be difficult to measure for most events for two reasons.
First, most events detected by \textit{Roman} will be too faint to observe from the ground, particularly with small telescopes.
Second, for events without caustic-crossings, the separation between the Earth and \textit{Roman}'s orbit at L2 will not be large enough to reveal a microlensing parallax measurement.
 Fortunately, for events with caustic-crossings the Earth-L2 separation yields a useful microlensing parallax measurement as \citet{wyrzykowski2020} demonstrate.
Moreover, for events with anomalies due to terrestrial planets, the microlens parallax may be measurable even in the absence of caustic-crossing features \citep{gould2003}.
 
Recently, planetary systems composed of low-mass dwarfs and a giant planet beyond the snow line of the dwarfs have been routinely detected from the KMTNet microlensing survey, even though most of masses of the host stars were estimated from a Bayesian analysis.
OGLE-2018-BLG-1428 is one such planetary system.
In this paper, we present the analysis of the planetary event OGLE-2018-BLG-1428, which has a short-duration caustic-crossing anomaly.
Although the finite source effect was measured from the caustic-crossing feature, the microlens parallax was not measured.
Therefore, the physical parameters of the lens system are estimated from a Bayesian analysis.

\section{observation} 

The planetary lensing event OGLE-2018-BLG-1428 is located at equatorial coordinates $({\rm RA}, {\rm decl.})_{\rm J2000}$ = (17:42:11.69, $-26$:08:16.4), corresponding to the Galactic coordinates $(l, b) = (1.99, 2.11)$.
The event was first alerted at August 6 2018 by the Optical Gravitational lensing Experiment (OGLE; \citealt{udalski2015}).
OGLE uses 1.3 m Warsaw telescope with $1.4\, \rm deg^{2}$ field of view (FOV) at the La Campanas Observatory in Chile. 
The event lies in the OGLE-IV field BLG652 with a low cadence of $\Gamma \simeq 0.01-0.1\, {\rm hr}^{-1}$.
In addition, the event is very near the edge of the OGLE chip and therefore has many missing data points due to small pointing variations.
In spite of this fact, OGLE alerted it at HJD-245000 (${\rm HJD}^\prime$)=8337.32, just before the peak.
However, due to the sparseness of the data points, the short-duration ($\sim 1$ day) caustic-crossing anomaly was not covered.

In 2018, KMTNet started to run its own alert system, but only for the northern bulge fields \citep{kim2018b}.
From the KMTNet alert system, OGLE-2018-BLG-1428 was independently announced at ${\rm HJD}^\prime$=8337.68, and it was designated as KMT-2018-BLG-0423.
KMTNet uses three identical telescopes with $4\, \rm deg^{2}$ FOV, which are individually located at CTIO in Chile (KMTC), SAAO in South Africa (KMTS), and SSO in Australia (KMTA).
The event lies in the KMT field BLG18 with cadence of $\Gamma \simeq 1$ hr$^{-1}$. 
With this cadence, the anomaly was well covered by KMTNet.
While most of KMTNet data were taken in the $I$ band, some of them were taken in the $V$ band in order to characterize the source star.
However, we found that the extinction toward the event, $A_{\rm I}=3.07$, is high, and thus it is difficult to use the $V$ band data  to constrain the source color.
To estimate the source color $(I-H)$, we used the $H$-band data of the VVV microlensing survey \citep{navarro2017, navarro2018}, which will be described in Section 4.
The KMTNet data were reduced by pySIS based on Difference Image Analysis (DIA; \citet{tomaney1996, alard1998, albrow2009}).

\section{Light Curve Analysis}
\subsection{Standard Model}

OGLE-2018-BLG-1428 is a binary lensing event with a clear caustic-crossing anomaly, which lasts $\sim 1\, {\rm day}$.
In order to describe a standard binary lensing event, seven lensing parameters are needed.
They include three single lensing parameters $(t_0,u_0,\te)$, three binary lensing parameters $(s,q,\alpha)$, and the source radius normalized to the angular Einstein radius of the lens $\thetae$ ($\rho=\thetas/\thetae$).
Here, $t_0$ is the peak time of the event, $u_0$ is the separation (in units of $\thetae$) between the lens and the source at $t_0$, $\te$ is the crossing time of the Einstein radius, $s$ is the star-planet separation in units of $\thetae$, $q$ is the planet-star mass ratio, and $\alpha$ is the angle between the source trajectory and the binary axis.
In the binary lensing modeling process, the observed fluxes of each observatory at a given time $t$ are modeled as $F_{i}(t) = A_{i}(t)f_{s,i} + f_{b,i}$, where $A_i$ is the magnification at the $i$th observatory and $f_{s,i}$ and $f_{b,i}$ are the source and blended fluxes at the $i$th observatory, respectively.
The $(f_{s,i},f_{b,i})$ are obtained from a linear fit.

In order to find the best-fit solution, we conduct a grid search over $(s,q,\alpha)$, which have the ranges of $-1 \leqslant {\rm log} s \leqslant 1$, $-4 \leqslant {\rm  log} q < 0 $, and $0 \leqslant \alpha < 2\pi$, respectively.
During the grid search, the $(s,q)$ are fixed, and the other parameters $(t_0,u_0,\te,\alpha,\rho)$ are allowed to vary in a Markov Chain Monte Carlo (MCMC) chain.
From the grid search, we find three local solutions including binary and planetary lens models with $(s,q,\alpha) = (1.29,0.0095,1.74),\ (0.85,0.0126,1.82)$, and $(1.35,0.0015,1.74)$.
We then conduct additional modeling in which the local solutions are set to the initial values and all parameters are allowed to vary.
As a result, we find that the best-fit solution of the event is the planetary lens model with $(s,q)=(1.42,0.0017)$, not the binary lens model.
The planetary lens model is favored by $\delcs=493$ relative to the binary lens model.
In this case, there is no $s \leftrightarrow 1/s$ degeneracy.
Figure 1 shows the light curve of the best-fit planetary lens model.
The best-fit lensing parameters are listed in Table 1.

Because the source crosses the caustic, we should consider the limb darkening of the finite source star in the modeling.
Considering the source type (discussed in Section 4), we assume that the source has solar metallicity, effective temperature $T_{\rm eff}=4750\, \rm K$, surface gravity ${\rm log} g \simeq 3.0$, and microturbulent velocity $v_{t}=2.0\, \rm km\, s^{-1}$.
 We thus adopt the limb darkening coefficient $\Gamma _{I}=0.51$ \citep{claret2000} and use Equation (7) of \citet{chung2019} for the source brightness profile.

\subsection{Investigation of Microlens Parallax}

Because the event timescale of $\te=24\, \rm days$ is relatively short and the source is relatively faint ($I_s = 19$), we do not expect to be able to measure the parallax.
However, we attempt to do so for completeness.
The orbital motion of the lens system can mimic the microlens parallax signal (\citealt{batista2011}, \citealt{skowron2011}).
We thus model the event adding both the microlens parallax and lens orbital motion.
The microlens parallax is described by $\bpie = (\pien,\piee)$, while the lens orbital motion is described by $ds/dt$ and $d\alpha/dt$, which are the instantaneous changes of the binary separation and the orientation of the binary axis, respectively.
From this, we find that although the parallax+orbital model is improved by $\delcs = 123$ relative to the standard model, it yields a weirdly high parallax magnitude $\pie \sim 5.5$.

In order to identify the source of the $\chi^{2}$ improvement and check for systematics, we build the cumulative distribution of $\delcs$ between the two models as a function of time.
As shown in Figure 2, the $\chi^{2}$ improvement comes from KMTC and KMTS (especially the former) while there is essentially no improvement for OGLE and KMTA.
We thus check the systematics of KMT data by binning them to $1$ per day.
These investigations show that the $\delcs$ improvement primarily comes from structures in the KMTC and KMTS data that are not seen in KMTA or OGLE.
Thus, they are likely due to correlated noise rather than a real signal.

Hence, we conduct the parallax+orbital remodeling with partial data sets for KMTC and KMTS and full data sets for KMTA and OGLE.
We restrict KMTC and KMTS data to data taken over the anomaly, i.e., in the range $8330.0 < {\rm HJD^{\prime}}<8342$.
The result shows that the $\delcs$ between the standard and the parallax+orbital models is $16$.
No orbital motion, $(ds/dt,d\alpha/dt) = (0,0)$ is within $3\sigma$ of the best-fit values, meaning these parameters are not significantly detected.
By contrast, $(\pien,\piee)=(0,0)$ is more than $3\sigma$ from the best-fit values, implying that there could be some real signal due to parallax.
However, this does not mean the parallax has to be large.
The contours are broad, allowing for a wide range of parallax values.
For example, the parallax values of $(\pien,\piee) = (0.4,-0.2)$ are compatible with the data at $\delcs = 6$. See Figure 3.
Since these values are not unreasonable, the parallax could be real.

\subsection{Xallarap Effect}
However, the parallax-like effects could be due to xallarap (source orbital motion).
We thus check the xallarap model.
Figure 4 shows the $\chi^2$ distribution for the best-fit xallarap solutions as a function of a fixed binary source orbital period $P$.
If the estimated parallax is real, the best-fit xallarap solution should appear at $P = 1.0\,\rm yr$, because the parallax is caused by the orbital motion of Earth.
As shown in Figure 4, the best-fit xallarap solution is at $P=0.2\,\rm yr$, not $P=1.0\,\rm  yr$, but several other solutions including $P=1.0\,\rm  yr$ have $\chi^2$ near the best solution.
The $\delcs$ between the best-fit parallax and xallarap solutions is $\delcs = 34$.
This suggests that the parallax solution is wrong and the large (and so, suspicious) parallax value is actually due to xallarap effects or systematics in the data.

We first check that all xallarap solutions are physically reasonable.
For each xallarap solution, we have two key parameters: the orbital period of binary source motion $P$ and the counterpart of the parallax $\xi_{\rm E}$.
The $\xi_{\rm E}$ is defined as $\xi_{\rm E}=a_{\rm s}/\hat{r}_{\rm E}$, where $a_{\rm s}$ is the semimajor axis of the source and $\hat{r}_{\rm E}$ is the Einstein radius projected to the source plane.
Thus, the source semimajor axis is
\begin{equation}
a_{\rm s} = \xi_{\rm E} \hat{r}_{\rm E}\, , \ \ \hat{r}_{\rm E} /{\rm AU} = \thetae\ds.
\end{equation}
As discussed in Section 4, the source is a G-type giant in the bulge, $\thetae = 0.377\, \rm mas$, and we assume $\ds=8.0\, \rm kpc$.
The source mass is thus $\sim 1M_\odot$, and then $\hat{r}_{\rm E} = 3.02\, \rm AU$.

According to Kepler's third law,
\begin{equation}
{M_{\rm tot}\over{M_\odot}}\left(P\over{\rm yr}\right)^{2} = {\left(a_{\rm tot}\over{\rm AU}\right)^{3}},
\end{equation}
where $M_{\rm tot} = M_{\rm s} + M_{\rm comp}$ and $a_{\rm s}/a_{\rm tot} = M_{\rm comp}/M_{\rm tot}$.
Here $M_{\rm s}$ and $M_{\rm comp}$ are the masses of the source and source companion, respectively.
We can parameterize Equation (2) by $Q=M_{\rm comp}/M_{\rm s}$.
Equation (2) then becomes
\begin{equation}
{(1+Q)^2\over{Q^3}} = {M_{\rm s}\over{M_\odot}} {(P/{\rm yr})^{2}\over{a_{\rm s}^{3}}}.  
\end{equation}
Using the estimated $\hat{r}_{\rm E}$, it is
\begin{equation}
{(1+Q)^2\over{Q^3}} = 0.036 {(P/{\rm yr})^{2}\over{\xi_{\rm E}^{3}}}.
\end{equation}
We solve this cubic equation for $Q$.
Then if $0.1 < Q < 1.0$, the solution is ``physically reasonable".
That is, the companion will be a ``typical main-sequence star".

The results for each $P$ and $\xi_{\rm E}$ are $Q=2.1,\ 10.1,\ 333.7,\ 2309.1,\ 5051.8,\ 30294.9,\ 53230.2,\ 497978.1,$ and $1456806.3$ for $P=0.1,\ 0.2,\ 0.4, 0.6,\ 0.8,\ 1.0,\ 1.2,\ 1.6$, and $2.0$, respectively. 
This means that the companion would be a very massive black hole.
Since the xallarap solution is not ``physically reasonable", the xallarap ``signal" is certainly not due to real xallarap.
Hence, it is due to systematics.
This investigation provides further evidence that systematics cause the parallax ``signal".
Therefore, we cannot measure the microlens parallax in this event (as was already anticipated due to its short Einstein timescale, $\te=24\ \rm days$).
As a result, we need a Bayesian analysis to estimate physical parameters of the lens system, which will be discussed in Section 5.

\section{Angular source radius}

KMTNet data were taken in the $I$- and $V$-bands in order to measure the instrumental source color.
Usually, the source color $(V-I)$ is measured from the linear regression of the $V$ on $I$ flux.
However, because there is only one $V$ point that is sufficiently magnified to give a significant signal due to high extinction $A_{I} = 3.07$, we cannot measure a reliable $(V-I)$.

In order to determine a reliable source color, we use the VVV $H$-band catalog.
Because the source is bright and the best-fit model shows negligible blending ($f_s \gg f_b$), we can attempt to measure the offset between the baseline object and the clump in the instrumental color magnitude diagram (CMD).
Here, we assume the baseline object corresponds to the source star due to negligible blending.

Figure 5 shows the calibrated $(I-H,I)$ and $(V-I,I)$ CMDs.
The CMDs have been calibrated by first applying the OGLE-IV calibration constants to the OGLE-IV data and then transforming the instrumental KMTC pyDIA data to the calibrated OGLE-IV system, in which the KMTC data are already matched to the VVV stars before the calibration.
From the $(I-H,I)$ CMD, we find that $(I-H,I)_{\rm cl} = (3.68, 17.55)$ and $(I-H,I)_{\rm s} = (3.52, 18.97)$, thus $\Delta (I-H)=-0.15$.
Using \citet{bessell1988}, we find that this corresponds to $\Delta (V-I)=-0.11$.
From the $(V-I,I)$ CMD, we find that $(V-I,I)_{\rm cl}=(3.63,17.57)$ and $(V-I,I)_{\rm s}=(3.37,18.97)$, thus $\Delta (V-I)=-0.26$.
Note that this baseline measurement derives from a stacking of several dozen images and so is more reliable than the regression method, which relies on a single magnified $V$ point.
We finally adopt that $\Delta (V-I)=-0.19$ by taking the average of these two values.
The instrumental source magnitude is $I=18.87$ from the best-fit model, and the magnitude and flux of the calibrated source is $I=18.99$ and $f_{\rm s}=0.4017$.
The source angular radius $\thetas$ is estimated from the intrinsic color and magnitude of the source, in which are determined from
\begin{equation}
(V-I,I)_{0} = (V-I,I)_{{\rm cl},0} + \Delta (V-I,I).
\end{equation}
With the measured $\Delta (V-I)=-0.19$, $I = 18.99$, and $(V-I, I)_{\rm cl,0}=(1.06,14.37)$ (\citealt{bensby2011}, \citealt{nataf2013}), we find that $(V-I,I)_0 = (0.87, 15.76)$, indicating that the source is a G-type giant.
We then estimate the source angular radius by using the $VIK$ color-color relation of \citet{bessell1988} and the color-surface brightness relation of \citet{kervella2004}.
From this, it is found that $\thetas = 2.717\pm 0.164\ \mu as$.
With the $\thetas$ and $\rho$, the Einstein angular radius of the lens is determined by
\begin{equation}
\thetae = \thetas/\rho = 0.373 \pm \ 0.026\  \rm mas 
\end{equation}
and the relative lens-source proper motion is
\begin{equation}
\mu_{\rm rel} = \thetae/\te = 5.58 \pm\ 0.38\  \rm mas\ yr^{-1}. 
\end{equation}

\section{Lens properties}

Because the parallax measurement is unreliable, we perform a Bayesian analysis to estimate physical properties of the lens, i.e., the mass and distance.
The Bayesian analysis implicitly assumes that all stars have an equal probability to host a planet of the measured mass ratio.
The Bayesian analysis is carried out with the same procedures as \citet{jung2018} did, but we use a new Galactic model based on more recent data and scientific understanding.
The new Galactic model includes the bulge mean velocity and dispersions taken from \textit{Gaia}, disk density profile  and disk velocity dispersion from the Robin-based model in \citet{bennett2014}, while the bulge mean velocity is generally zero and the bulge density profile is the same as the one in \citet{jung2018}.
However, we know the proper motion of the source $(\mu_\alpha, \mu_\delta)=(-2.607 \pm 2.371,-3.014 \pm 1.751)$ from \textit{Gaia}, even though its error is big.
We thus use the proper motion value as the mean velocity of the source for bulge-bulge events.

In addition, we should consider the extinction at a given distance for the lens brightness.
For the extinction to the lens $A_{\rm L}$, we use the following equation \citep{bennett2015,batista2015}
\begin{equation}
A_{i,\rm L} = {1-e^{-|D_{\rm L}/(h_{\rm dust} {\rm sin}\ b)|}\over{1-e^{-|D_{\rm S}/(h_{\rm dust} {\rm sin}\ b)|}}}\ A_{i,\rm S},
\end{equation}
where the index $i$ denotes the passband: $V$, $I$, or $K$, and the dust scale height is $h_{\rm dust} = 120\ \rm pc$.
Here we adopt the extinction to the source of $A_{I,\rm S} = 2.98$ and $A_{K,\rm S}=0.35$ from the VVV/KMTC CMD analysis and $VIK$ color-color relation of \citet{bessell1988}, which were discussed in Section 4.

Figure 6 shows the results of the Bayesian analysis.
From this, we find that the lens is a sub Jupiter-mass planet $M_{\rm p}=0.77^{+0.77}_{-0.53} \ M_{\rm J}$ orbiting a star $M_{\rm h}=0.43^{+0.33}_{-0.22} \, M_{\odot}$ at a distance $D_{\rm L}=6.22^{+1.03}_{-1.51}\, \rm kpc$, and the projected star-planet separation is $3.30^{+0.59}_{-0.83}\, \rm AU$.$  $
This indicates that OGLE-2018-BLG-1428L is likely to be an M dwarf star hosting a sub Jupiter-mass planet beyond the snow line based on $a_{\rm snow}=2.7\ (M/M_{\odot})$ \citep{kennedy2008}.
However, it could be also a K or a G dwarf.
The lens distribution in Figure 6 shows that the lens is located in the disk and bulge with equal probability.
This is consistent with the relative proper motion of $5.6\ \rm mas\ yr^{-1}$.

Figure 6 also shows the Bayesian distributions for the brightness of host star.
The distributions show that if the host star is a main-sequence star, its brightness is $I_L = 25.3^{+0.9}_{-2.4}$ and $K_L = 20.7^{+0.7}_{-1.8}$.
Considering the brightness of the giant source star with $K=14.7$, the lens star is $\sim 240$ times fainter than the source.
This high contrast between the source and the lens makes it difficult to resolve the two stars by follow-up observations.
However, for both MOA-2007-BLG-400 \citep{bhattacharya2020} and MOA-2013-BLG-220 \citep{vandorou2020}, the lens mass measured from Keck is much closer to the $2\sigma$ upper limit from the Bayesian analysis than the median.
Thus, considering the $2\sigma$ upper limit of the lens brightness, the lens with $K=17.1$ is 9 times fainter than the source.
Recently, \citet{bhattacharya2020} reported that lens star with $K=18.9$, which is $\sim 10$ times fainter than the source and is $\sim 50\, \rm mas$ away from the source, can be detected at a separation of $0.53$ FWHM with Keck.
For this event, because the proper motion is $5.6\, \rm mas\ yr^{-1}$, the lens will be separated from the source by $56\, \rm mas$ in 2028.
Hence, it seems plausible that a lens at $K \sim 17$ would be detectable by Keck, while for a lens at $K \sim 21$ it would be hard to detect with Keck.
If the lens is a very faint star at $K \sim 21$, the lens can be resolved by a 30m telescope equipped with a state of the art laser guide star adaptive optics system, even though the contrast between source and lens is high.
Such a measurement can resolve the nature of the lens and confirm the results of the Bayesian analysis.

\section{Summary}

We analyzed the event OGLE-2018-BLG-1428 with a caustic-crossing feature.
From the Bayesian analysis, it is found that the lens is a star $M_{\rm L}=0.43^{+0.33}_{-0.22} \ M_{\odot}$ hosting a sub Jupiter-mass planet $M_{\rm p}=0.77^{+0.77}_{-0.53} \ M_{\rm J}$, at a distance $D_{\rm L}=6.22^{+1.03}_{-1.51}\ {\rm kpc}$, and the projected separation between the star and the planet is $3.30^{+0.59}_{-0.83}\, {\rm AU}$, suggesting that the planet orbits beyond the snow line of the host. 
The lens distance distribution and the proper motion $\mu_{\rm rel} = 5.6\  \rm mas\ yr^{-1}$ indicate that the lens is located in the disk and bulge with equal probability.
The lens can be resolved by adaptive optics of a 30m telescope in the future.

\section*{Acknowledgements}
Work by Y.-H. Kim and S.-J. Chung was supported by the KASI (Korea Astronomy and Space Science Institute) grant 2021-1-830-08. Work by A.G. was supported by JPL grant 1500811. Work by C.H. was supported by the grant of National Research Foundation of Korea (2019R1A2C2085965 and 2020R1A4A2002885). The OGLE project has received funding from the National Science Centre, Poland, grant MAESTRO 2014/14/A/ST9/00121 to A.U. This research has made use of the KMTNet system operated by the KASI and the data were obtained at three sites of CTIO in Chile, SAAO in South Africa, and SSO in Australia. 

\section*{Data Availability}
The data underlying this article will be shared on reasonable request to the corresponding author.


\begin{table*}
\begin{minipage}{126mm}
\caption{Best-fit lensing parameters.}
\label{tab-one}
\begin{tabular}{lc}
\hline\hline
 Parameter         &                   \\
\hline
$\chi^2$/dof                     &   $2902.14/2927 $       \\        
$t_0$ (HJD$^\prime$)     & $8339.6157 \pm 0.0950$  \\
$u_0$                               &  $0.7002 \pm 0.0030$     \\ 
$\te$ (days)                     &  $24.4448 \pm 0.1858$  \\
$s$                                   &  $1.4233 \pm 0.0019$    \\
$q(10^{-3})$                   &  $1.7144\pm 0.0553$      \\  
$\alpha$ (rad)                &  $1.7271 \pm 0.0052$    \\     
$\rho$                            &  $0.0073 \pm 0.0002$   \\    
$f_{s,\rm kmt}$            &  $0.4480 \pm 0.0022$     \\    
$f_{b,\rm kmt}$            &  $-0.0485 \pm 0.0021$   \\     
$f_{s,\rm ogle}$            &  $0.3901 \pm 0.0028$    \\    
$f_{b,\rm ogle}$            &  $0.0088 \pm 0.0028$    \\    
\hline
\end{tabular}

\medskip
Note- HJD$^\prime$ = HJD - 2450000.
\end{minipage}
\end{table*}

\begin{table*}
\begin{minipage}{126mm}
\caption{Physical lens parameters.}
\label{tab-two}
\begin{tabular}{lc}
\hline\hline
 Parameter                       &               \\
\hline   
$M_{\rm host}$ $(M_\odot)$         &    $0.43^{+0.33}_{-0.22}$  \\
$M_{\rm p}\, (M_{\rm J})$              &    $0.77^{+0.77}_{-0.53}$  \\
$D_{\rm L}$ (kpc)                           &    $6.22^{+1.03}_{-1.51}$    \\ 
$a_\perp$ (au)                                &    $3.30^{+0.59}_{-0.83}$   \\
$\thetae$                                          &    $0.373 \pm 0.026$             \\
$\mu_{\rm rel}\, (\rm mas\ yr^{-1})$    &   $5.58 \pm 0.38$       \\           
\hline
\end{tabular}
\end{minipage}
\end{table*}


\begin{figure*}
\centering
\includegraphics[width=150mm]{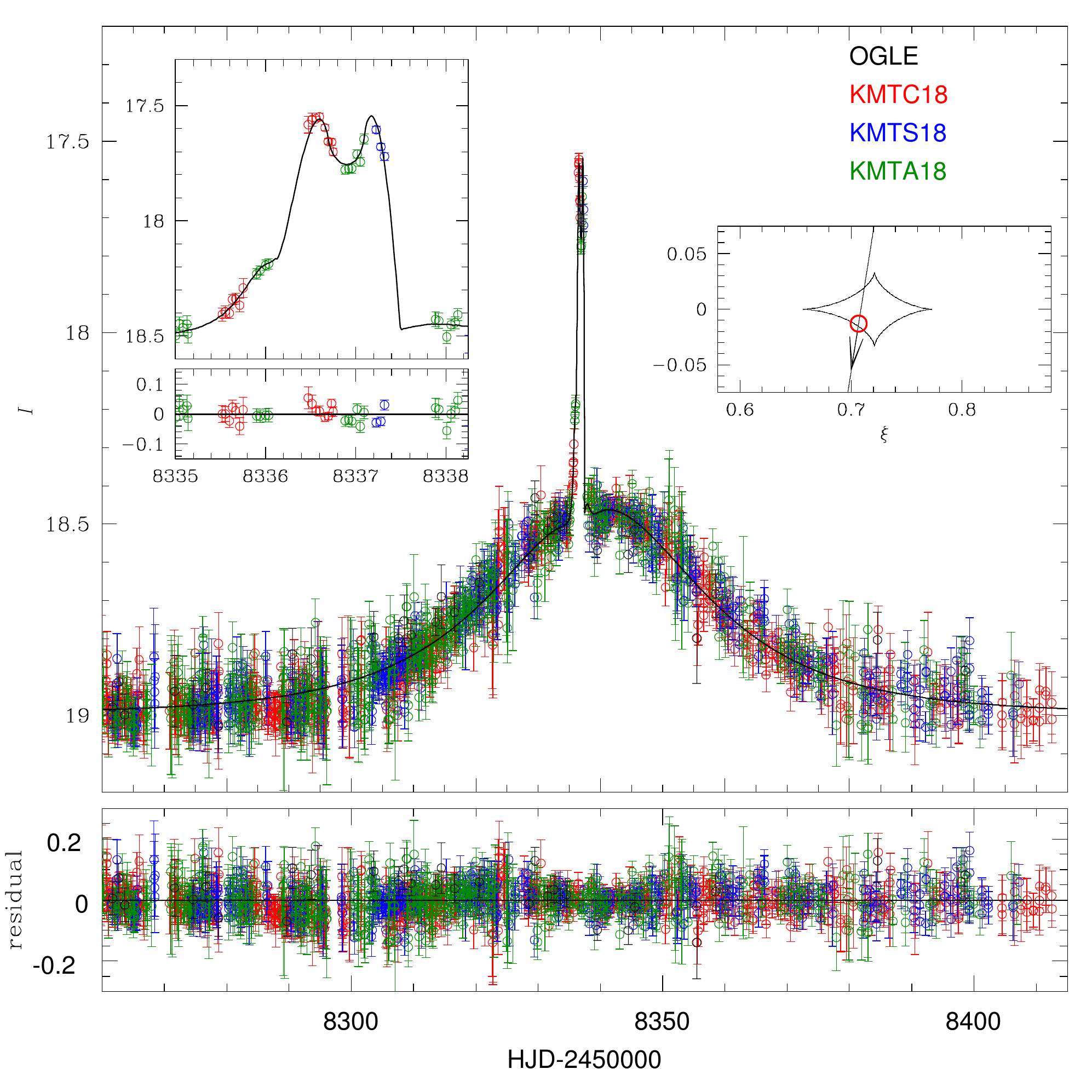}
\caption{Lightcurve of the best-fit lensing model.
The right inset shows the source trajectory crossing the planetary caustic.\label{fig:f1}}
\end{figure*}

\begin{figure*}
\centering
\includegraphics[width=150mm]{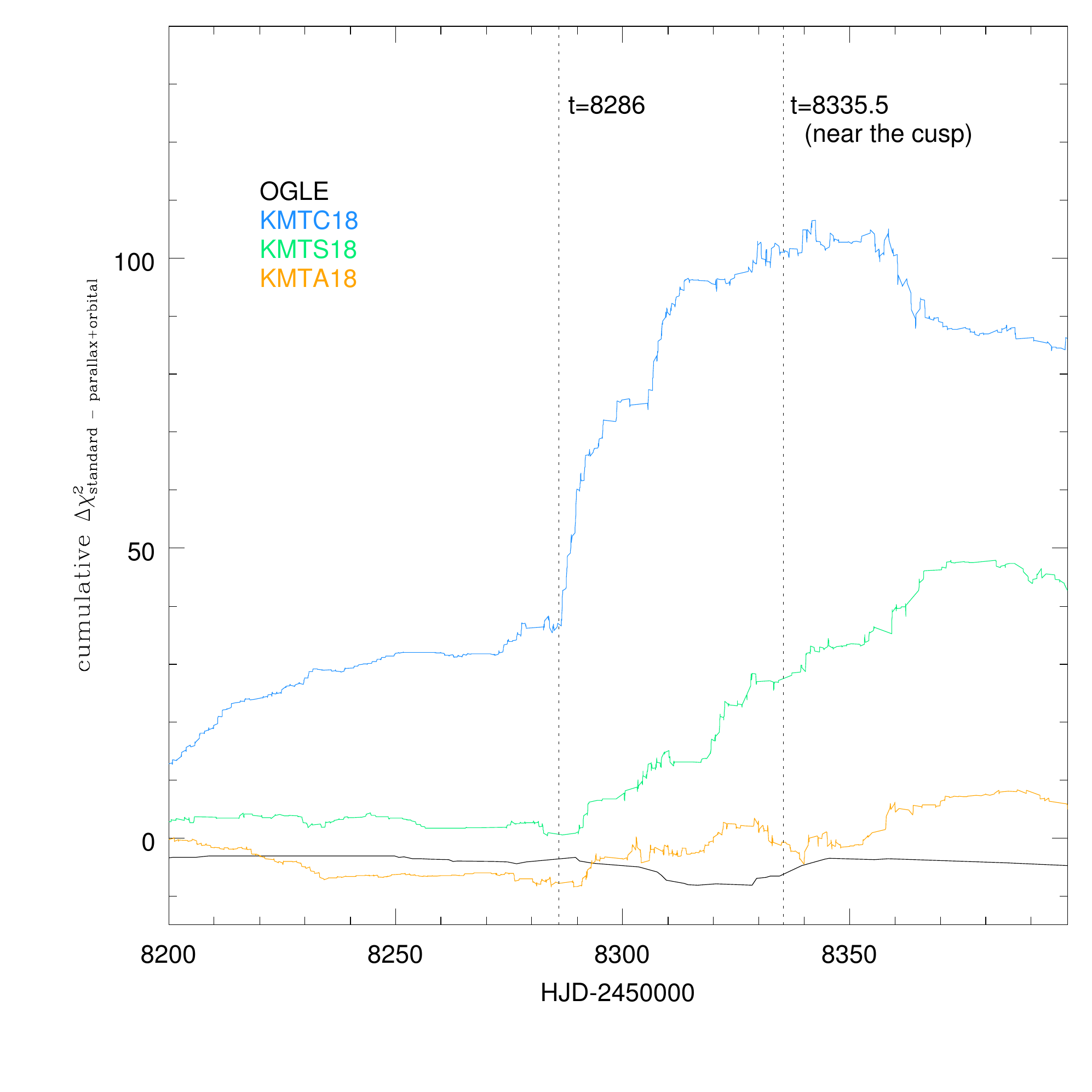}
\caption{Cumulative $\delcs$ between the standard and the parallax+orbital models. This shows that the $\chi^2$ improvement for the parallax+orbital model comes from KMTC and KMTS data sets.\label{fig:f2}}
\end{figure*}

\begin{figure*}
\centering
\includegraphics[width=150mm]{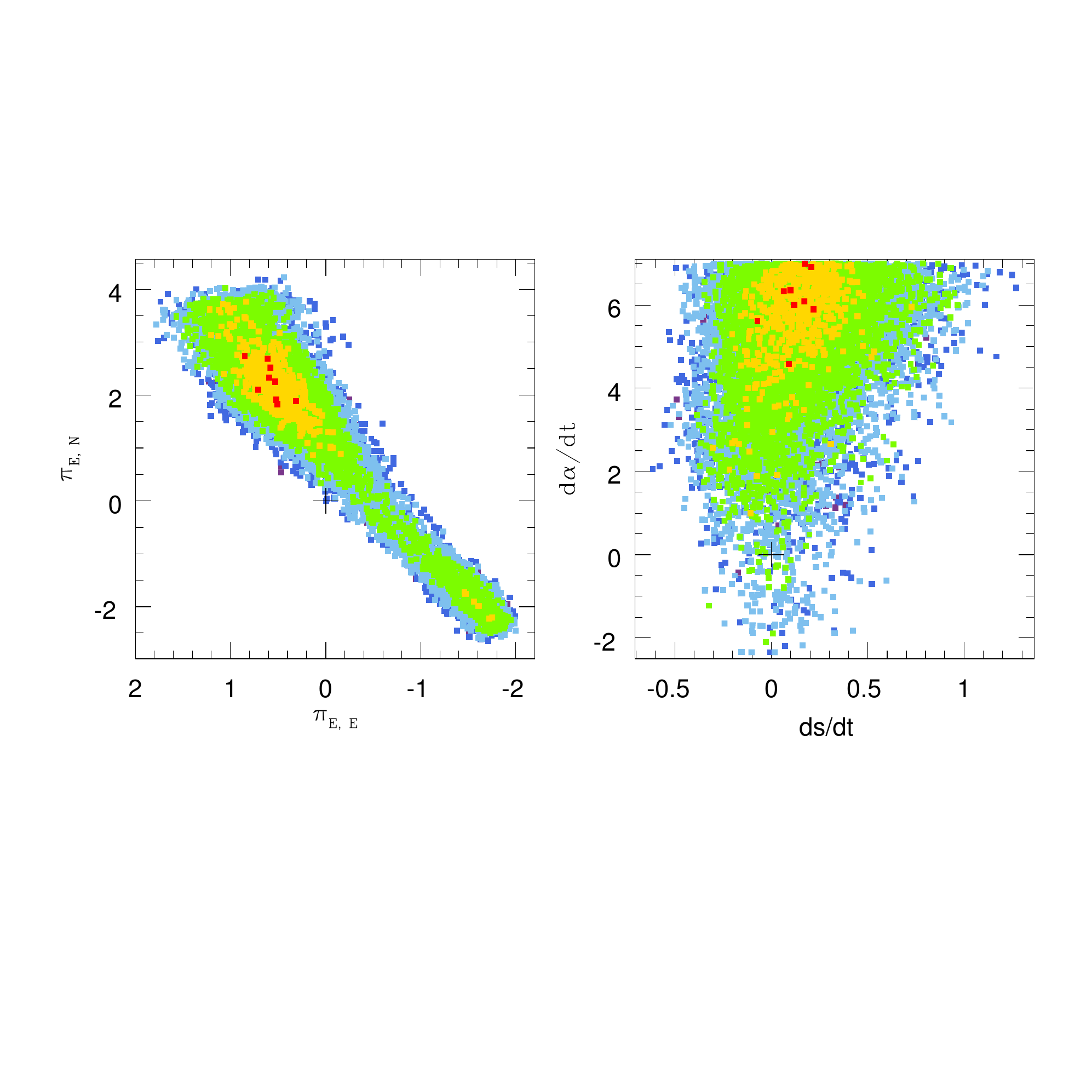}
\caption{$\chi^2$ distributions of the parallax+orbital model with partial data sets of the anomaly range (restricted to the anomaly) for KMTC and KMTS and full data sets for KMTA and OGLE.\label{fig:f3}}
\end{figure*}

\begin{figure*}
\centering
\includegraphics[width=150mm]{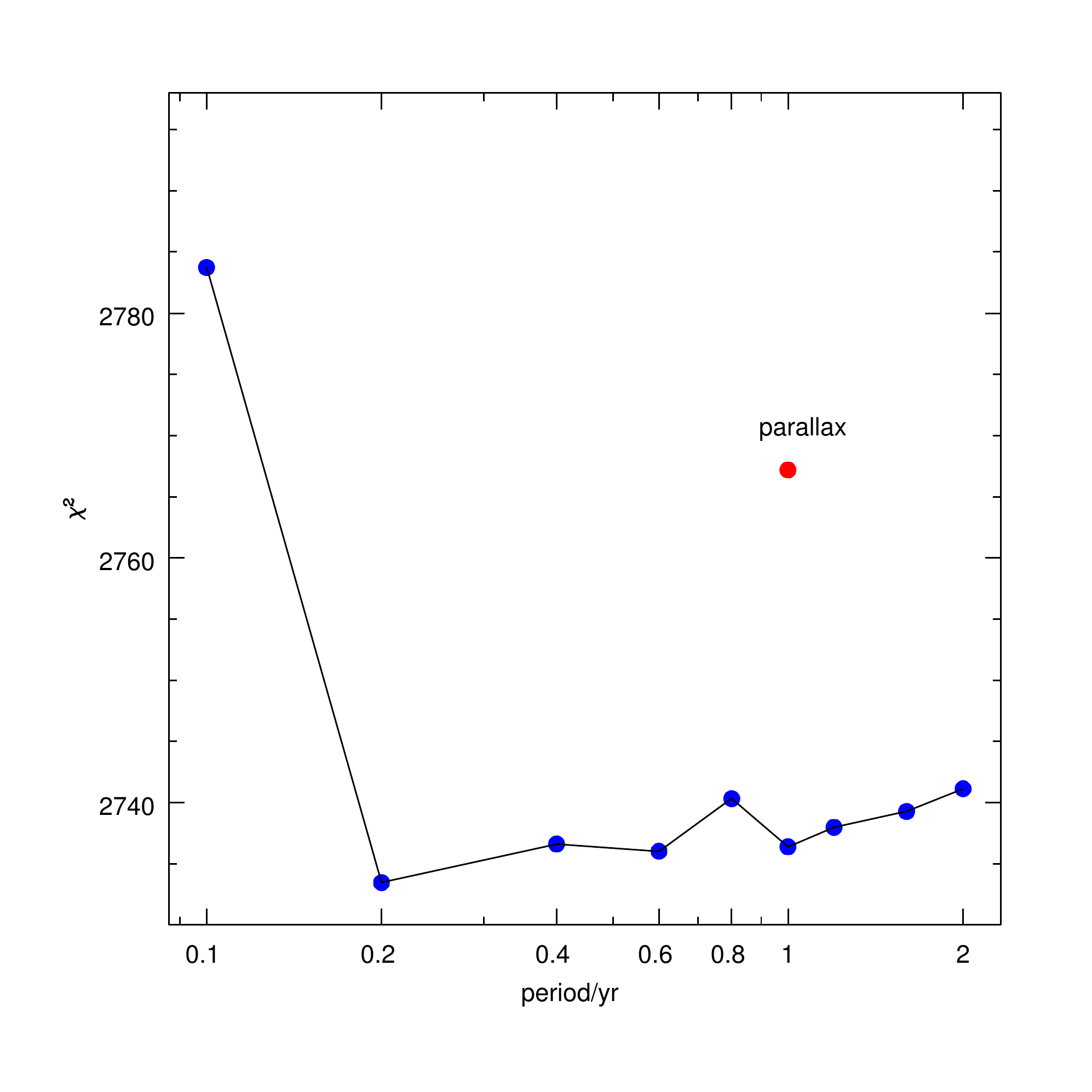}
\caption{$\chi^2$ distribution for the best-fit xallarap solutions as a function of a fixed binary source orbital period $P$.
The red dot is the $\chi^2$ of the best-fit parallax+orbital model.\label{fig:f4}}
\end{figure*}

\begin{figure*}
\centering
\includegraphics[width=150mm]{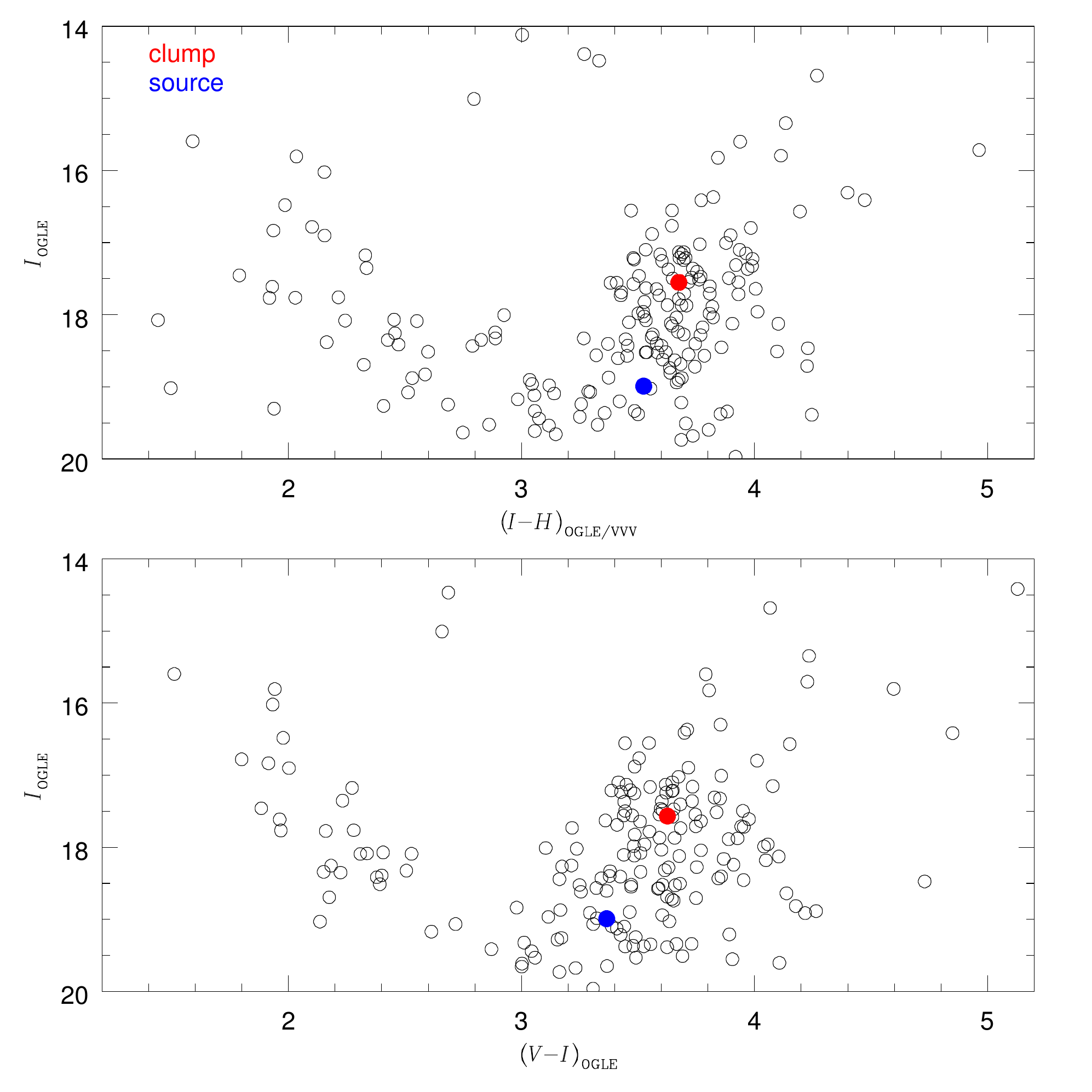}
\caption{Calibrated $(I-H, I)$ and $(V-I, I)$ color-magnitude diagrams (CMDs).
The red and blue dots are the red clump giant centroid and source position.\label{fig:f5}}
\end{figure*}

\begin{figure*}
\centering
\includegraphics[width=150mm]{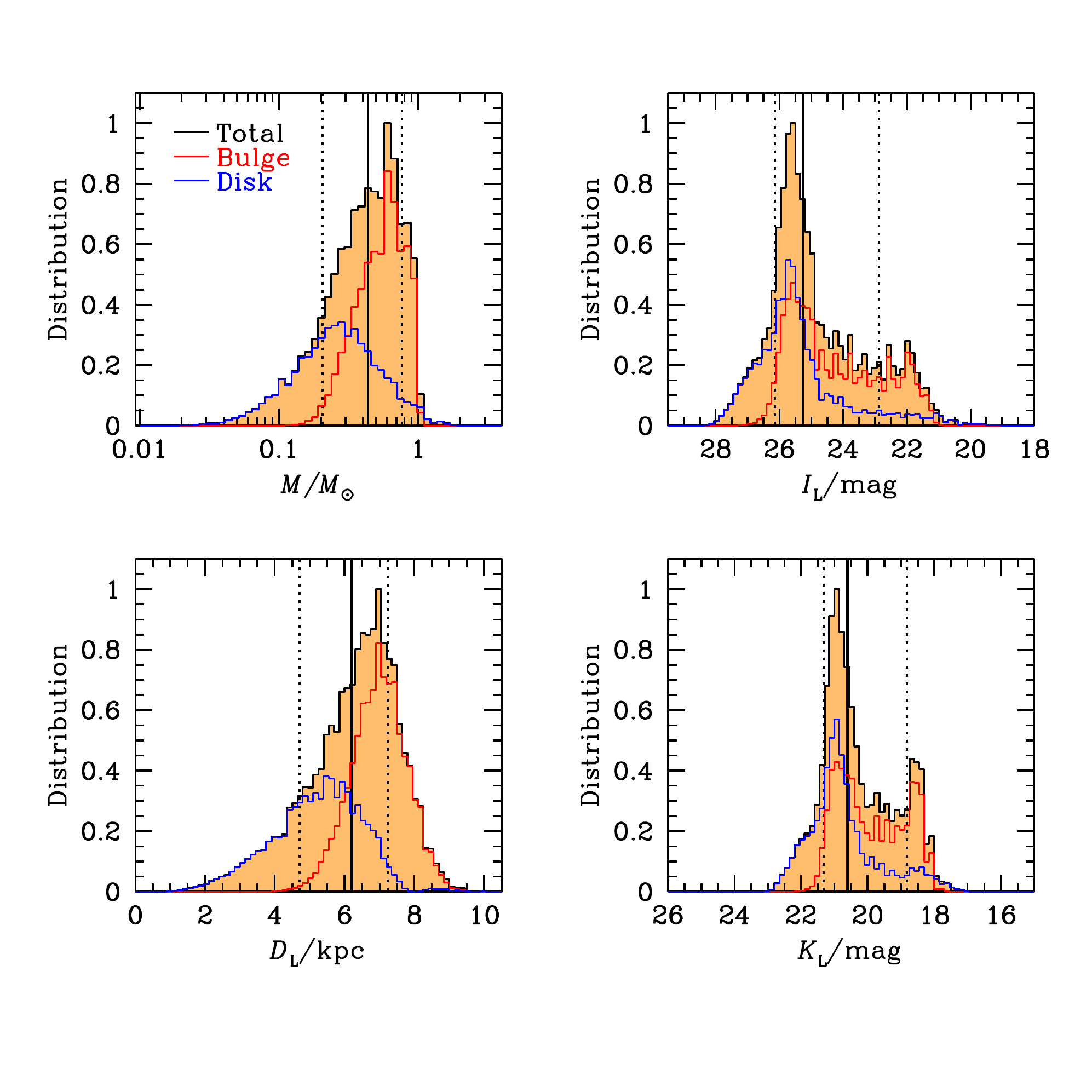}
\caption{Bayesian distributions for physical parameters of the host star.
The vertical solid line indicates the median value, while the two vertical dotted lines indicate the confidence intervals of $68\%$.
 \label{fig:f6}}
\end{figure*}

\label{lastpage}
\end{document}